\documentclass[11pt]{article}
\usepackage{amsmath,amsthm,amssymb,epic,eepic}
\newcommand{\nc}{\newcommand}
\nc{\mref}[1]{(\ref{#1})}
\nc{\vt}{v_{2\gL_0}}
\nc{\vo}{v_{\gL_0}}
\nc{\vot}{v_{\gL_1+\gL_0}}
\nc{\vw}{v_{\gL_1}}
\nc{\ppmm}{\genfrac{}{}{-10pt}{10pt}{++}{--}}
\nc{\wom}[5]{\Omega\left(\left.\begin{array}{ll}{#1}&{#2}\\{#3}&{#4}\end{array}\right|{#5}\right)}
\nc{\com}[5]{\chi\left(\left.\begin{array}{ll}{#1}&{#2}\\{#3}&{#4}\end{array}\right|{#5}\right)}
\nc{\we}[5]{W\left(\left.\begin{array}{ll}{#1}&{#2}\\{#3}&{#4}\end{array}\right|{#5}\right)}
\nc{\ce}[6]{C^{#6}\left(\left.\begin{array}{ll}{#1}&{#2}\\{#3}&{#4}\end{array}\right|{#5}\right)}
\nc{\lmat}[6]{\ell_{#6}\left(\left.\begin{array}{ll}{#1}&{#2}\\{#3}&{#4}\end{array}\right|{#5}\right)}
\nc{\lmats}[5]{L\left(\left.\begin{array}{ll}{#1}&{#2}\\{#3}&{#4}\end{array}\right|{#5}\right)}
\nc{\hmat}[6]{h_{#6}\left(\left.\begin{array}{ll}{#1}&{#2}\\{#3}&{#4}\end{array}\right|{#5}\right)}
\nc{\hmats}[5]{H\left(\left.\begin{array}{ll}{#1}&{#2}\\{#3}&{#4}\end{array}\right|{#5}\right)}
\nc{\web}[5]{\overline{W}\left(\left.\begin{array}{ll}{#1}&{#2}\\{#3}&{#4}\end{array}\right|{#5}\right)}
\nc{\wep}[5]{W'\left(\left.\begin{array}{ll}{#1}&{#2}\\{#3}&{#4}\end{array}\right|{#5}\right)}
\nc{\wes}[5]{W^*\left(\left.\begin{array}{ll}{#1}&{#2}\\{#3}&{#4}\end{array}\right|{#5}\right)}
\nc{\wess}[5]{W^{**}\left(\left.\begin{array}{ll}{#1}&{#2}\\{#3}&{#4}\end{array}\right|{#5}\right)}
\nc{\cet}[7]{C^{#6}_{#7}\left(\left.\begin{array}{ll}{#1}&{#2}\\{#3}&{#4}\end{array}\right|{#5}\right)}
\nc{\bcet}[7]{\bar{C}^{#6}_{#7}\left(\left.\begin{array}{ll}{#1}&{#2}\\{#3}&{#4}\end{array}\right|{#5}\right)}
\nc{\wet}[7]{W^{#6}_{#7}\left(\left.\begin{array}{ll}{#1}&{#2}\\{#3}&{#4}\end{array}\right|{#5}\right)}
\nc{\bwet}[7]{\overline{W}^{#6}_{#7}\left(\left.\begin{array}{ll}{#1}&{#2}\\{#3}&{#4}\end{array}\right|{#5}\right)}
\nc{\wec}[7]{\widetilde{W}^{#6}_{#7}\left(\left.\begin{array}{ll}{#1}&{#2}\\{#3}&{#4}\end{array}\right|{#5}\right)}
\nc{\wgen}[6]{W^{#6}\left(\left.\begin{array}{ll}{#1}&{#2}\\{#3}&{#4}\end{array}\right|{#5}\right)}
\nc{\wgenp}[6]{W^{*{#6}}\left(\left.\begin{array}{ll}{#1}&{#2}\\{#3}&{#4}\end{array}\right|{#5}\right)}
\nc{\wo}[5]{\Omega\left(\left.\begin{array}{ll}{#1}&{#2}\\{#3}&{#4}\end{array}\right|{#5}\right)}
\nc{\wsgen}[8]{{#8}^{#6}_{#7}\left(\left.\begin{array}{ll}{#1}&{#2}\\{#3}&{#4}\end{array}\right|{#5}\right)}
\nc{\qbinom}[2]{{\genfrac{[}{]}{0pt}{}{{#1}}{{#2}}}_{q}}
\nc{\hhg}[4]{\phi\left({{{#1}\,\,\,{#2}}\atop{{#3}}};
                     {#4}\right)}
\nc{\fullhhg}[5]{{_1}\phi_2\left({{{#1}\,\,\,{#2}}\atop{{#3}}};
                     {#4},{#5}\right)}
\nc{\bra}[1]{\langle #1 |}
\nc{\ket}[1]{| #1 \rangle}
\nc{\qp}[2]{({#1}\, ; \, {#2})_{\infty}}
\nc{\qpf}[1]{({#1}\, ; \, q^4)_{\infty}}
\nc{\pp}[1]{({#1}\, ; \, p)_{\infty}}
\nc{\qpp}[1]{({#1}\, ; \, p, q^4)_{\infty}}
\nc{\sect}{\section}
\nc{\ssect}{\subsection}
\nc{\sssect}{\subsubsection}
\nc{\ud}[1]{\underline{{#1}}}
\nc{\isomo}{\buildrel {\sim} \over \longrightarrow}
\nc{\Aff}{\operatorname{Aff}}
\nc{\ot}{\otimes}
\nc{\er}{\end{array}}
\nc{\bev}[1]{\begin{equation}\begin{array}{#1}}
\nc{\eeq}{\end{equation}}
\nc{\be}{\begin{eqnarray}}
\nc{\ee}{\end{eqnarray}}
\nc{\ben}{\begin{eqnarray*}}
\nc{\een}{\end{eqnarray*}}
\nc{\bec}{\begin{equation}\begin{array}{lll}}
\nc{\eec}{\end{array}\end{equation}}
\nc{\ed}{\end{document}}
\nc{\half}{\ensuremath{\frac{1}{2}}}
\nc{\Hom}{\operatorname{Hom}}
\nc{\End}{\operatorname{End}}
\nc{\vac}{|\textrm{vac}\rangle}
\nc{\tvac}{|\widetilde{\textrm{vac}}\rangle}
\nc{\dvac}{\langle\textrm{vac}}
\nc{\dtvac}{\langle\widetilde{\textrm{vac}}}
\nc{\id}{\operatorname{id}}
\nc{\ra}{\rightarrow}  
\nc{\lra}{\longrightarrow}
\nc{\uqp}{U^{\prime}_q (\widehat{sl}_2)}
\nc{\uqbp}{U_q (b_+)}
\nc{\uqbm}{U_q (b_-)}
\nc{\ub}{U^{\prime}_q (b_+)}
\nc{\vsl}{V(\sigma(\lambda))}
\nc{\vl}{V(\lambda)}  
\nc{\bu}{\bullet}
\nc{\an}{{\ell}}
\nc{\slth}{\widehat{\mathfrak{sl}}_2\hskip 1pt}
\newcommand{\uq}{U_q\bigl(\slth\bigr)}
\nc{\ws}{\;\;}
\nc{\qu}{{1\ov 4}}
\nc{\hif}{\hb{ if }}
\nc{\hev}{\hb{ is even }}
\nc{\hod}{\hb{ is odd }}
\nc{\Tr}{{\rm Tr}}
\nc{\ad}{{\rm Ad}}
\nc{\hb}{\hbox}
\nc{\nn}{\nonumber} 
\nc{\curlra}{\buildrel{\sim}\over\longrightarrow}
\nc{\epp}{\varepsilon^{\prime}} 
\nc{\ol}{\overline}
\nc{\pl}{\prod\limits} 
\nc{\sli}{\sum\limits} 
\nc{\nin}{\noindent}

\nc{\ga}{\alpha}
\nc{\gb}{\beta}
\nc{\gd}{\delta}
\nc{\gep}{\varepsilon}
\nc{\gz}{\zeta}
\nc{\gt}{\theta}
\nc{\gk}{\kappa}
\nc{\gl}{\lambda}
\nc{\gp}{\phi}
\nc{\gs}{\sigma}
\nc{\go}{\omega}
\nc{\gn}{\nu}
\nc{\gr}{\rho}

\nc{\s}{\sigma}
\nc{\ep}{\varepsilon}
\nc{\z}{\zeta}
\nc{\g}{\gamma}
\nc{\zi}{\zeta^{-1}}

\nc{\gG}{\Gamma}
\nc{\gD}{\Delta}
\nc{\gT}{\Theta}
\nc{\gL}{\Lambda}
\nc{\gO}{\Omega}
\nc{\gP}{\Phi}

\nc{\cF}{\mathcal{F}}
\nc{\cP}{\mathcal{P}}
\nc{\cS}{\mathcal{S}}
\nc{\cN}{\mathcal{N}}
\nc{\cD}{\mathcal{D}}
\nc{\cH}{\mathcal{H}}
\nc{\cO}{\mathcal{O}}
\nc{\cT}{\mathcal{T}}
\nc{\cQ}{\mathcal{Q}}
\nc{\cW}{\mathcal{W}}
\nc{\cR}{\mathcal{R}}

\nc{\C}{\mathbb{C}}
\nc{\Q}{\mathbb{Q}}
\nc{\R}{\mathbb{R}}
\nc{\Z}{\mathbb{Z}}
\nc{\N}{\mathbb{N}}

\nc{\fg}{\mathfrak{g}}

\nc{\bi}{\bar{i}}
\nc{\bj}{\bar{j}}
\nc{\bgr}{\bar{\rho}}
\nc{\bA}{\bar{\alpha}}
\nc{\bB}{\bar{\beta}}
\nc{\bC}{\bar{\gamma}}
\nc{\by}{\bar{y}}

\nc{\tf}{\tilde{f}}
\nc{\te}{\tilde{e}}
\nc{\ts}{\tilde{s}}
\nc{\tgP}{\widetilde{\Phi}}
\nc{\tgPs}{\tilde{\Psi}}
\nc{\tgn}{\tilde{\nu}}
\nc{\tgl}{\tilde{\lambda}}
\nc{\tge}{\tilde{\eta}}
\nc{\txi}{\tilde{\xi}}
\nc{\tep}{\tilde{\epsilon}}

\nc{\cB}{\check{b}}

\nc{\goto}{\mapsto}
\nc{\embed}{\hookrightarrow}
\nc{\rien}{\emptyset}
\nc{\lb}[1]{\label{#1}}
\nc{\Nt}{\frac{N}{2}}
\nc{\vn}{\hspace*{-33truemm}}
\nc{\vm}{\hspace*{-0truemm}}
\nc{\ti}{t^{-1}}
\nc{\vb}{v^{(1)}}
\nc{\vbn}{v^{(n)}}
\nc{\us}{\underline{s}}
\nc{\up}{\underline{p}}
\nc{\bp}{\bar{p}}
\nc{\bpi}{\bar{p}^{(i)}}
\nc{\bpip}{\bar{p}^{(i+1)}}
\nc{\vz}{V^{(1)}_z}
\nc{\vzn}{V^{(n)}_z}
\nc{\vzo}{V^{(1)}_1}
\nc{\piz}{\pi_z^{(1)}}
\nc{\pizn}{\pi_z^{(n)}}
\nc{\pis}{\pi_{(z,\us)}}
\nc{\bW}{\overline{W}}
\nc{\bQ}{\overline{Q}}
\nc{\tQ}{\widetilde{Q}}
\nc{\bT}{\overline{T}}
\nc{\note}[1]{\vspace*{-5mm}\marginpar[left]{\scriptsize\bf{#1}}}
\nc{\eqdef}{:=}
\nc{\lu}{^{(\lambda)}}
\nc{\vone}{v_{\Lambda_1}}
\nc{\vzero}{v_{\Lambda_0}}
\nc{\ds}{\displaystyle}
\begin{document}
\bibliographystyle{unsrt}
\begin{center}
{\Large \bf The Entanglement Entropy of Solvable Lattice Models\\[10mm] }
{\large \bf Robert Weston}\\[6mm]
{\it Department of Mathematics, Heriot-Watt University,\\
Edinburgh EH14 4AS, UK. R.A.Weston@ma.hw.ac.uk}\\[5mm]
January 2006 (revised Feb 2006)\\[10mm]
\end{center}
\begin{abstract}
\noindent 
We consider the spin $\gk/2$ analogue of the XXZ
quantum spin chain. We compute the entanglement entropy $S$
associated with splitting the infinite chain into two semi-infinite
pieces. In the scaling limit, we find 
$ S\simeq \frac{c_\gk}{6} \ln(\xi)+\ln(g)+\cdots$.  
Here $\xi$ is the correlation length 
and $c_\gk=\frac{3\gk}{\gk+2}$ is the central charge associated with the 
$\slth$ WZW model at level $\gk$. $\ln(g)$ is the boundary entropy of the
WZW model. Our result extends previous 
observations and suggests that this is a simple and perhaps 
rather general
way both of extracting the central charge of the ultraviolet CFT 
associated with the scaling limit of a solvable lattice model, and
of matching lattice and CFT boundary conditions.
\end{abstract}

\vspace*{10mm}
\nin In this letter, we consider the spin $\gk/2$ analogue of the XXZ
quantum spin chain (which corresponds to the choice $\gk=1$). We derive
an exact formula for the entanglement entropy $S^{(i,\gk)}$ associated with splitting the
infinite chain into two semi-infinite pieces. We consider the scaling
limit and find that

\be S^{(i,\gk)}\simeq\frac{c_\gk}{6} \ln(\xi)+ \ln(g^{(i,\gk)})+
{\cal C}_\gk.\lb{Sorig}\ee
The first term involves the correlation length $\xi$ 
and $c_\gk=\frac{3\gk}{\gk+2}$ which is the central charge of the 
$\slth$ WZW model at level $\gk$. The scaling limit of our quantum
spin chain is known to correspond to
a massive perturbation of this conformal field theory \cite{FMS}. 
This leading term involves the central charge
in the way predicted in \cite{CC04} and generalizes the $\gk=1$
result derived there. 
The second term in \mref{Sorig} is the $\slth$ level
$\kappa$ WZW model boundary entropy of Affleck and
Ludwig \cite{aflu91}. It depends on the 
boundary conditions of our lattice model labelled by $i\in\{0,1,\cdots,\gk\}$.
The third term, ${\cal C}_\gk$,
gives another finite,  $\gk$-dependent contribution.
We derive \mref{Sorig} 
directly from the known expression for the partition
function in terms of $\uq$
characters.

The definition and meaning of entanglement entropy can be found in many places
in the literature, see, for example, \cite{BBSS96}.
Our starting point is the
density matrix defined by $\rho=\vac \dvac| \in \End(\cH)$, where $\vac\in \cH$ is 
the lowest energy eigenstate in the infinite tensor product space $\cH$
 of our quantum spin chain. The space $\cH$ is naturally a tensor
product of semi-infinite left and right spaces $\cH=\cH_l\ot \cH_r$.
We are interested in the reduced density matrix 
defined as the partial trace 
\ben
\rho_l=\Tr_{\cH_r}( \rho)\in
\End(\cH_l).\een
The entanglement entropy we consider is then defined by
\be S=-\Tr_{\cH_l} (\rho_l \ln\rho_l).\lb{entS}\ee

In \cite{Bax82}, Baxter has discussed a range of 2D solvable lattice models
which are related to 1D quantum spin chains via a transfer matrix, and
whose partition functions can be written in the form 
\ben Z(x)=\Tr_{\cH_l}(x^{2H_{CTM}}),\een
where $H_{CTM}$, the corner-transfer-matrix Hamiltonian, acts on the left-hand space $\cH_l$
 associated with the
corresponding quantum spin chain, and has eigenvalues $0,1,2,\cdots$. 
It was shown in \cite{Nish95}, that for such models it is possible to
identify the reduced density matrix
$\rho_l$ discussed above as
\ben \rho_l=\frac{x^{2 H_{CTM}}}{Z(x)}.\een
A simple argument leading to this result is presented in 
\cite{PKL98}.
The entanglement entropy defined by \mref{entS} is given by 
\ben S=-\Tr_{\cH_{l}}(\rho_l \ln(\rho_l))
=-\frac{d}{dn} \Tr_{\cH_{l}}(\rho_l^n)\big|_{n=1},\een
which we can now write as
\be
S=-\frac{d}{dn}\left(\frac{Z(x^n)}{Z(x)^n}\right)\Big|_{n=1}=
\ln(Z(x))-x \ln(x)\, \frac{Z'(x)}{Z(x)}.\label{Se}\ee
This approach was used in \cite{CC04} in order to compute the
leading contribution to the entanglement entropy for the Ising and XXZ models.
In this paper we generalize these results to higher-spin models.

The models of interest to us are the 
spin $\gk/2$ analogues of the 6-vertex model. These are
solvable 2D statistical-mechanical models defined in terms 
of spin variables $\ep\in\{0,1,\cdots,\gk\}$ that live on the
edges of a square lattice. The Boltzmann weight of a spin configuration
is given as the product of local weights $R(\gz_1/\gz_2)^{\ep_1,\ep_2}_{\ep'_1,\ep'_2}$
associated with the following configuration around each vertex:

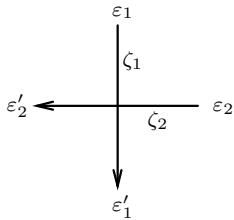
\setlength{\unitlength}{0.00035in}
\begingroup\makeatletter\ifx\SetFigFont\undefined%
\gdef\SetFigFont#1#2#3#4#5{%
  \reset@font\fontsize{#1}{#2pt}%
  \fontfamily{#3}\fontseries{#4}\fontshape{#5}%
  \selectfont}%
\fi\endgroup%
{\renewcommand{\dashlinestretch}{30}
\begin{picture}(3905,3105)(-5000,200)
\thicklines
\path(2250,2775)(2250,375)
\path(2190.000,615.000)(2250.000,375.000)(2310.000,615.000)
\path(1290.000,1635.000)(1050.000,1575.000)(1290.000,1515.000)
\path(1050,1575)(3450,1575)
\put(2175,2925){\makebox(0,0)[lb]{\smash{{{\SetFigFont{8}{9.6}{\rmdefault}{\mddefault}{\updefault}$\ep_1$}}}}}
\put(3675,1500){\makebox(0,0)[lb]{\smash{{{\SetFigFont{8}{9.6}{\rmdefault}{\mddefault}{\updefault}$\ep_2$}}}}}
\put(2175,0){\makebox(0,0)[lb]{\smash{{{\SetFigFont{8}{9.6}{\rmdefault}{\mddefault}{\updefault}$\ep'_1$}}}}}
\put(600,1500){\makebox(0,0)[lb]{\smash{{{\SetFigFont{8}{9.6}{\rmdefault}{\mddefault}{\updefault}$\ep'_2$}}}}}
\put(2325,2175){\makebox(0,0)[lb]{\smash{{{\SetFigFont{8}{9.6}{\rmdefault}{\mddefault}{\updefault}$\gz_1$}}}}}
\put(2700,1275){\makebox(0,0)[lb]{\smash{{{\SetFigFont{8}{9.6}{\rmdefault}{\mddefault}{\updefault}$\gz_2$}}}}}
\end{picture}
}

\vspace*{2mm}

The $R$-matrix $R(\gz)$ is understood in modern algebraic language to
be an intertwiner of spin $\gk/2$ evaluation representations of the algebra
$\uq$. For explicit formulae for $R(\gz)$ and a full discussion of this
algebraic picture see \cite{JM,idzal93}. In the simplest $\gk=1$ case, $R(\gz)$
gives the weights of the 6-vertex model. $\gk=2$ corresponds to the 19-vertex model,
etc. 

In this paper, we do not require explicit formulae for the Boltzmann weights,
but just make use of known results for the partition function $Z(x)$ (here
$x=-q$, where $q$ is the deformation parameter of the $\uq$ vertex model). 
However, we are dealing with the infinite-volume model and {\it do} need to specify 
boundary conditions. We consider the model in the antiferromagnetic regime
corresponding to the choice $0<x<1$, and consider a partition function,
denoted $Z^{(i,\gk)}(x)$, which is defined to be the 
weighted sum over all spin configurations which 
are fixed at some finite but arbitrarily large distance from a nominal
centre of our lattice to the following ground state configuration
labelled by $i\in\{0,1,\cdots,\gk\}$ (with $\bar{i}:=\gk-i$):

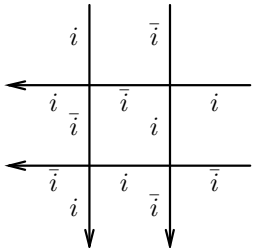
\setlength{\unitlength}{0.00035in}
\begingroup\makeatletter\ifx\SetFigFont\undefined%
\gdef\SetFigFont#1#2#3#4#5{%
  \reset@font\fontsize{#1}{#2pt}%
  \fontfamily{#3}\fontseries{#4}\fontshape{#5}%
  \selectfont}%
\fi\endgroup%
{\renewcommand{\dashlinestretch}{30}
\begin{picture}(3644,3659)(-6000,-10)
\put(2122,1672){\makebox(0,0)[lb]{\smash{{{\SetFigFont{10}{9.6}{\rmdefault}{\mddefault}{\updefault}$i$}}}}}
\put(922,1672){\makebox(0,0)[lb]{\smash{{{\SetFigFont{10}{9.6}{\rmdefault}{\mddefault}{\updefault}$\bar{i}$}}}}}
\put(922,3022){\makebox(0,0)[lb]{\smash{{{\SetFigFont{10}{9.6}{\rmdefault}{\mddefault}{\updefault}$i$}}}}}
\put(2122,3022){\makebox(0,0)[lb]{\smash{{{\SetFigFont{10}{9.6}{\rmdefault}{\mddefault}{\updefault}$\bar{i}$}}}}}
\put(2122,472){\makebox(0,0)[lb]{\smash{{{\SetFigFont{10}{9.6}{\rmdefault}{\mddefault}{\updefault}$\bar{i}$}}}}}
\put(922,472){\makebox(0,0)[lb]{\smash{{{\SetFigFont{10}{9.6}{\rmdefault}{\mddefault}{\updefault}$i
          $}}}}}
\put(622,840){\makebox(0,0)[lb]{\smash{{{\SetFigFont{10}{9.6}{\rmdefault}{\mddefault}{\updefault}$\bar{i}$}}}}}
\put(1672,840){\makebox(0,0)[lb]{\smash{{{\SetFigFont{10}{9.6}{\rmdefault}{\mddefault}{\updefault}$i$}}}}}
\put(3022,840){\makebox(0,0)[lb]{\smash{{{\SetFigFont{10}{9.6}{\rmdefault}{\mddefault}{\updefault}$\bar{i}$}}}}}
\put(622,2040){\makebox(0,0)[lb]{\smash{{{\SetFigFont{10}{9.6}{\rmdefault}{\mddefault}{\updefault}$i$}}}}}
\put(1672,2040){\makebox(0,0)[lb]{\smash{{{\SetFigFont{10}{9.6}{\rmdefault}{\mddefault}{\updefault}$\bar{i}$}}}}}
\put(3022,2040){\makebox(0,0)[lb]{\smash{{{\SetFigFont{10}{9.6}{\rmdefault}{\mddefault}{\updefault}$i$}}}}}
\thicklines
\path(2422,3622)(2422,22)
\path(2362.000,262.000)(2422.000,22.000)(2482.000,262.000)
\path(262.000,1282.000)(22.000,1222.000)(262.000,1162.000)
\path(22,1222)(3622,1222)
\path(262.000,2482.000)(22.000,2422.000)(262.000,2362.000)
\path(22,2422)(3622,2422)
\path(1222,3622)(1222,22)
\path(1162.000,262.000)(1222.000,22.000)(1282.000,262.000)
\end{picture}
}

The partition function $Z^{(i,\gk)}(x)$
is given in terms of the 
principally specialised $\uq$ character associated with 
the level $\gk$ irreducible highest-weight
representation $V(\gl_i)$ with weight $\gl_i=i\gL_1+(\gk-i)\gL_0$ (where $\gL_1$ and
$\gL_0$ are the fundamental $\uq$ weights) \cite{DJKMO89}. The explicit formula
\cite{KaP84} is
\be
Z^{(i,\gk)}(x)&=&x^i\,\, \Tr_{V(\gl_i)}(x^{-2\rho})=
\frac
{\theta_{x^{2(\gk+2)}} (x^{2(i+1)}) }
{\theta_{x^{4}} (x^{2}) }.\lb{char}
\ee
Here, $\rho=\gL_1+\gL_0$ is the $\uq$ principal derivation (see \cite{idzal93}), and we use the
standard notation
 \be
\theta_w(z)=(z;w)_{\infty} \ (w\, z^{-1};w)_{\infty} \ (w;w)_{\infty},\quad
\hb{and}\ws  (z;w)_{\infty}=\sli_{n=0}^{\infty}(1-z\,w^n).
\ee
A useful identity is 
\be
(x^2;x^{2})_{\infty}=(x^2;x^{2m})_{\infty} (x^4;x^{2m})_{\infty}
 \cdots (x^{2m};x^{2m})_{\infty}.
\lb{id2}\ee
As an example, consider the $\gk=1$ case. After using \mref{id2}, we
obtain
$Z^{(i,1)}(x)=1/(x^2;x^4)_{\infty}$. 
Note, that $Z^{(0,1)}(x)=Z^{(1,1)}(x)$ in this case, reflecting the
$\ep=0\leftrightarrow \ep=1$ symmetry of the 6-vertex model.

In this paper, we are interested in the entanglement entropy defined 
by 
\be S^{(i,\gk)}=\ln(Z^{(i,\gk)}(x)) - x \ln(x)
\frac{Z^{(i,\gk)\,'}(x)}{Z^{(i,\gk)}(x)}.\lb{ent1}\ee
However, before computing this object, let us first discuss the scaling
limit
of the vertex models.
The correlation length
of the $\gk=1$ model is given \cite{JKM73,Bax82} by 
\be \xi^{-1} =- \half\ln\left( 
\frac{1-k'}{1+k'}
\right),\lb{xi}\ee
where $k'$ is the conjugate modulus associated with elliptic nome
$x$ (a gentle introduction to elliptic functions 
can be found in Chapter 15 of \cite{Bax82}).
If we define $\epsilon$ by $x=e^{-\epsilon}$, then the  $\xi \ra \infty$ 
scaling limit corresponds to $\epsilon\ra 0$, $x\ra 1$. In this 
limit $k' \simeq 4 e^{-\pi^2/\epsilon} \ra 0$, and so  from \mref{xi} we have $\xi^{-1}\simeq k'$
such that 
\be \ln(\xi) \simeq \frac{\pi^2}{2\epsilon}.\lb{corrl}\ee
It follows from the results of \cite{Sogo84} that \mref{corrl} also
holds for general $\gk$.

The
series appearing if we substitute \mref{char} into \mref{ent1} with 
$x=e^{-\epsilon}$ converges slowly for small
$\epsilon$ and is  not appropriate for a consideration of the
scaling limit. To proceed, we Poisson re-sum the series.
Let us show how this works for a certain building block  $T(a,b)$
which appears in characters.
We define $T(a,b)$ for $0<a< b$ by 
\ben T(a,b):=(x^{a};x^b)_{\infty} (x^{b-a};x^b)_{\infty}.\een
The contribution to the entanglement entropy from such a building block is
\ben S(a,b)&=&\ln(T(a,b))-x \ln(x) \frac{dT(a,b)}{dx}  \\
&=& \sli_{n=0}^{\infty}
\Big(f(\epsilon(a+nb))+f(\epsilon(-a+(n+1)b))\Big)
= \sli_{n\in \Z} f(|\epsilon(a+nb)|),\\
\hb{where} \ws f(y)&:=&\ln(1-e^{-y}) +\frac{y}{1-e^y}.\een
We can use the Poisson summation formula (see p. 468 of \cite{Bax82})
to rewrite $S(a,b)$ as 
\ben S(a,b)&=& \frac{2}{\epsilon b} \sli_{n\in \Z} \widehat{f}\left(\frac{2\pi n}{\epsilon b}\right) 
e^{\frac{2\pi i n a}{b}},\quad
\hb{where} \ws \widehat{f}(y):=\int_{0}^{\infty}f(x) \cos(yx) dx = 
\frac{\pi}{2y} 
\left(\frac{\pi y-\sinh(\pi y)\cosh(\pi y)}{\sinh^2(\pi y)}\right).
\een
Noting that $\widehat{f}(0)=-\pi^2/3$, we arrive at the expression
\be S(a,b) = -\frac{2\pi^2}{3\epsilon b} +\frac{4}{\epsilon b}
 \sli_{n=1}^{\infty} 
\widehat{f}\left(\frac{2\pi n}{\epsilon b}\right) 
\cos\left(\frac{2\pi n a}{b}\right).\lb{mainseries}\ee
The first term clearly diverges as $\epsilon\ra 0$, whereas the 
series converges for $\epsilon$ small.

Using the fact that $\ds\lim_{\delta\ra 0_+}\frac{1}{\gd}\hat{f}(\gd)=-\pi/2$, it
follows that the limit of the series appearing in \mref{mainseries} is given by
\be
\lim_{\epsilon\ra 0_+} 
\frac{4}{\epsilon b} \sli_{n=1}^{\infty} 
\widehat{f}\left(\frac{2\pi n}{\epsilon b}\right) \cos\left(\frac{2\pi n a}{b}\right)
= - \sli_{n=1}^{\infty} \frac{1}{n} \cos\left(\frac{2\pi n a}{b}\right)=
\ln\left(\,2\sin\left(\frac{\pi a}{b}\right)\right).\ee
Hence, in the scaling limit we have 
\be S(a,b)\simeq  -\frac{2\pi^2}{3\epsilon b}
 +\ln\left(\,2\sin\left(\frac{\pi a}{b}\right)\right),\lb{Sscale}\ee
with other contributions vanishing as $\ep\ra 0_+$.

Now let us return to the computation of the entanglement entropy \mref{ent1}.
We can express $Z^{(i,\gk)}(x)$ entirely
in terms of elementary blocks $T(a,b)$ by making use of \mref{id2} and
the identification $(x^{b/2};x^b)_{\infty}=T(b/2,b)^{1/2}$.
We find
\bev{llll} Z^{(i,\gk)}(x)&=& 
\frac{T(2(i+1),2(\gk+2))}{T(2;4)^\half\, T(\gk+2,2(\gk+2))^\half \,T(2,2(\gk+2))\,
T(4,2(\gk+2)) \cdots T(\gk,2(\gk+2))} \quad &\hb{for}\ws \gk\ws \hb{even},\\[5mm]
Z^{(i,\gk)}(x)&=& 
\frac{T(2(i+1),2(\gk+2))}{T(2;4)^\half \, T(2,2(\gk+2))\,
T(4,2(\gk+2)) \cdots T(\gk+1,2(\gk+2))} \quad &\hb{for}\ws \gk\ws \hb{odd}.
\nn\eec
Each term $T(a,b)$ in the numerator will contribute $S(a,b)$ to the
entropy; each term $T(a,b)$ or $T(a,b)^\half$ in the
denominator
 will contribute
$-S(a,b)$ or $-\half S(a,b)$ respectively. 
Noting that all but one $T(a,b)$ terms have the same value of 
$b=2(\gk+2)$, and that $a$ only appears in \mref{mainseries} via the
cosine term,
it should be clear that a simple exact expression for $S^{(i,\gk)}$
then follows by using elementary combinatorics and trigonometric function identities.
In the scaling limit,
we use the form \mref{Sscale} and obtain
\be S^{(i,\gk)}\simeq \frac{\pi^2}{12 \ep}c_\gk+
\ln\left(\frac{2^\half \sin\left(\frac{\pi(i+1)}{(\gk+2)}
    \right)}{(\gk+2)^\half}\right),\quad \hb{for $\gk$ both odd and
  even,} 
\lb{epS}\ee
where $c_\gk=3\gk/(\gk+2)$.
 
The boundary contribution to the thermodynamic 
entropy of the $\slth$ level $\kappa$ WZW model was first
obtained in \cite{aflu91} and is given by $\ln(g^{(i,\gk)})$, where
$g^{(i,\gk)}$, the `ground state degeneracy', is 
\ben g^{(i,\gk)} = 
\frac{
\sin\left(\frac{\pi(i+1)}{(\gk+2)}\right)
 }
{ \sin\left(\frac{\pi}{(\gk+2)}\right)  }.\een
In the WZW context, the integer $i$ labels the Cardy state of the
boundary CFT \cite{Ca05}.  
It then follows that after making use of \mref{corrl} we arrive at expression
\mref{Sorig}
for $S^{(i,\gk)}$. The term ${\cal C}_{\gk}$ in \mref{Sorig} includes the 
extra terms in the $\ln$ function in \mref{epS} as well as $\epsilon$
independent, but possibly $\gk$ dependent, corrections to \mref{corrl}.

We would like to make several comments by way of conclusion:
\begin{itemize}
\item The leading term in \mref{Sorig} is universal and involves the central
charge in the way
discussed in detail in \cite{CC04,Cas04,Hol94}
(the scaling limit of the spin-$\gk/2$ model is known to correspond to a massive
perturbation of the $\slth$ level $\kappa$ 
WZW model with central charge $c_\gk$).

\item The second term in \mref{Sorig} depends upon the boundary
conditions
labelled by $i$. A similar contribution to the entanglement entropy
from each boundary 
was found in \cite{CC04} for critical model models on a finite 
length system with open boundaries (see equation (1.2) of
\cite{CC04} and try not to be confused by our reversion to the original $\ln(g)$ notation of \cite{aflu91} - see also \cite{ZBFS05}). It is an interesting and apparently new observation that
such a term is present for infinite size off-critical lattice models and survives in the
scaling limit.

\item It is possible to derive \mref{Sorig} by applying the 
conjugate modulus transformation $x=e^{-\epsilon}\ra
\tilde{x}=e^{-\pi^2/\epsilon}$ directly to
the 
theta functions appearing in equation \mref{char}. We have chosen to 
present a rather more pedestrian approach in the hope that it
may be applicable to a wider range of solvable lattice models, where
the theta function realization of characters is not so well
understood.

\item
Our approach to computing the entanglement entropy should be 
generalizable to other familiar examples such as RSOS models,
and hopefully to more exotic models. 
We hope that it might actually provide a simple alternative to 
the thermodynamic Bethe ansatz as a way of 
extracting the central charge of the ultraviolet CFT associated with 
a wide range of solvable lattice models. We also hope, given
the presence of the boundary entropy term, that 
this technique for computing entanglement entropy 
might generally be useful for matching physical off-critical
lattice boundary conditions with Cardy states in boundary CFT.

\end{itemize}

\subsubsection*{Acknowledgements}
The author would like to thank Vladimir Korepin for introducing him to this subject, and to
acknowledge funding from the European TMR network EUCLID (contract
number HPRN-CT-2002-00325). He would also like to thank the referees 
of the first version of this paper for pointing out that
our boundary term was equal to the boundary entropy given
in \cite{aflu91}

\baselineskip=13pt

\end{document}